\documentstyle[12pt,epsfig,axodraw]{article}
\begin{document}
\begin{flushright}
BIHEP-TH-2002-7
\end{flushright}

\newcommand{\Title}[1]{{\baselineskip=26pt \begin{center}
            \Large   \bf #1 \\ \ \\ \end{center}}}
\newcommand{\Author}{\begin{center}\large
       Medina Ablikim$^b$, Dong-Sheng Du$^{a,b}$, Mao-Zhi Yang$^{a,b}$\end{center}}
\newcommand{\Address}{\begin{center} \it
       $^a$ CCAST(World Laboratory), P.O.Box 8730, Beijing 100080, China\\

$^b$ Institute of High Energy Physics, Chinese Academy of
             Sciences, P.O.Box 918(4),\\ Beijing 100039, China

\end{center}}

\bigskip
\bigskip
\bigskip

\baselineskip=20pt

\Title{$D \to \pi\pi$ decays with Final State Interactions}
\Author

\Address
\vspace{2cm}

\begin{abstract}

We study $D \to \pi\pi$ decays with final state interactions considered
in one-particle-exchange method. A clear physical picture for final
state interactions based on quark and hadronic level diagrams is
presented. A strong phase is introduced for
hadronic effective couplings, which is crucial for explaining the
experimental data of $D^+\to \pi^+\pi^0$, $D^0\to \pi^+\pi^-$, and
$D^0\to \pi^0\pi^0$. Rescattering effects between
different $D$ decay channels are usually large. They are important
for obtaining correct branching ratios for $D\to \pi\pi$ decays
in theoretical calculation.

\end{abstract}

\section{Introduction}
\setcounter{equation}{0}

The study of heavy meson decays is important for understanding the
quark mixing sector of the standard model (SM). It is of great help for
determining the quark mixing parameters, searching for the sources
of $CP$ violation. However, the quarks in nature are not free,
they are bounded in hadrons by strong interactions which
are described by nonperturbative QCD. The strong interaction
permeates every processes. It almost masks all the information of
electroweak (EW) interaction. One must solve the problem of strong
interaction before making any meaningful measurements of EW or
quark flavor physics in experiment. Solving the problem of
nonperturbative QCD needs efforts in both experiment and theory.
In the near future BESIII and CLEO-c detector will provide high
precision data in charm physics including data on $D$ meson decays,
which will provide the possibility for understanding the physics
in charm sector and $D$ decays.

It is the high time to study $D$ meson two-body weak decays
beyond the factorization approach now \cite{BSW}. In general, if a process
happens in an energy scale where there are many resonance states, this process
must be seriously affected by these resonances \cite{fsi}. This is a highly
nonperturbtive effect. Near the scale of $D$ mass many resonance states
exist. $D$ meson decays must be affected by them seriously. After weak
decays the final state particles rescatter into other particle states
through nonperturbative strong interaction \cite{fsi,Donoghue}.
This is called final state
interaction (FSI). Every $D$ decay channels
can contribute to each other through final state interactions. One can
model this rescattering effect as one-particle exchange process
\cite{ope1,ope2}. That is to say that the final state particles be scattered
into other particle states by exchanging one resonance state
existing near the mass scale of $D$ meson.
There are also other ways to treat the nonperturbative and FSI effects in
nonleptonic $D$ decays. The readers
can refer to Ref.\cite{ref}

In this paper, we study $D^+\to \pi^+\pi^0$, $D^0\to \pi^+\pi^-$, and
$D^0\to \pi^0\pi^0$ decays. We use the one-particle-exchange method to
study the final state interactions in these decays. The magnitudes of
hadronic couplings needed here are extracted from experimental data on the measured
branching fractions of resonances decays. In addition, we introduce
a strong phase for the hadronic coupling which is
important for obtaining the correct branching ratios of $D\to\pi\pi$ decays.
The coupling constants
extracted from experimental data are small for $s$-channel contribution and
large for $t$-channel contribution. Therefore the $s$-channel contribution
is numerically negligible in $D\to\pi\pi$ decays. We can safely drop the
$s$-channel contribution in this paper.

The paper is organized as follows. Section II presents the calculation
in naive factorization approach. Section III gives the main scheme of
one-particle-exchange method. Section IV presents
the numerical calculation and discussions.
Finally a brief summary is given.

\section{Calculations in the factorization approach }
\setcounter{equation}{0}

We start with the low energy effective Hamiltonian for charm
decays \cite{buras}
\begin{equation}
  {\cal H} = \frac{G_F}{\sqrt2} \left[ \sum_{q=d,s} v_q \left(C_1Q_1^q +
         C_2Q_2^q \right)\right], \label{Hamiltonian}
\end{equation}
where $C_1$ and $C_2$ are the Wilson coefficients at $m_c$ scale,
$v_q$ is the product of Cabibbo-Kobayashi-Maskawa (CKM) matrix elements and
defined as
\begin{equation}
  v_q=V_{uq}V_{cq}^* ,
\end{equation}
and the current-current operators are given by
\begin{equation}
  Q_1^q=(\bar u q)_{V-A}(\bar q c)_{V-A}, \qquad
  Q_2^q=(\bar u_\alpha q_\beta)_{V-A}(\bar q_\beta c_\alpha)_{V-A}.
\end{equation}
We do not consider the contributions of the QCD and the
electroweak penguin operators in the decays of $D \to \pi\pi$,
since their effects are small in $D$ decays. The values of $C_1$
and $C_2$ at $m_c$ scale are taken to be \cite{buras}
$$ C_1=1.126,~~~~~C_2=-0.415 .$$

In the naive factorization approach, the decay amplitude can be
generally factorized into a product of two current matrix elements
and can be obtained from~(\ref{Hamiltonian})
\begin{eqnarray}
 && A(D^+ \to \pi^+\pi^0)=-\frac{G_F}{2}\;V_{ud}\;V_{cd}^*\;
      (a_1+a_2)\;if_{\pi}( m_D^2-m_{\pi}^2)\;F^{D\pi}(m_{\pi}^2) ,\nonumber\\
 && A(D^0 \to \pi^+\pi^-)=\frac{G_F}{\sqrt2} \; V_{ud}\;V_{cd}^*\;
      a_1\;if_{\pi}(m_D^2-m_{\pi}^2)\;F^{D\pi}(m_{\pi}^2) ,\\
 && A(D^0 \to \pi^0\pi^0)=-\frac{G_F}{2} \; V_{ud}\;V_{cd}^* \;
      a_2\;if_{\pi}(m_D^2-m_{\pi}^2)\;F^{D\pi}(m_{\pi}^2),\nonumber
\end{eqnarray}
where the parameters $a_1$ and $a_2$ are defined as
\begin{equation}
a_1=C_1+\frac{C_2}{N_c}, \qquad a_2=C_2+\frac{C_1}{N_c},
\end{equation}
with the color number $N_c=3$. The decay constant $f_{\pi}$ and the form factor $F^{D\pi}(0)$ take
the values of 0.133 GeV and 0.692 respectively. For $q^2$
dependence of the form factors, we take the BSW model \cite{BSW}, i.e., the
monopole dominance assumption:
\begin{equation}
F(q^2)=\frac{F(0)}{1-q^2/m_*^2},
\end{equation}
where $m_*$ is the relevant pole mass.

The decay width of a $D$ meson at rest decaying into $\pi\pi$ is
\begin{equation}
 \Gamma(D \to \pi\pi)=\frac{1}{8\pi}|A(D \to \pi\pi)|^2\frac{|\vec
p\;|}{m_D^2},
\end{equation}
where the momentum of the $\pi$ meson is given by
\begin{equation}
  |\vec p\;|=\frac{[m_D^2(m_D^2-4m_{\pi}^2)]^{1/2}}{2m_D}.
\end{equation}
The corresponding branching ratio is
\begin{equation}
 Br(D \to \pi\pi)=\frac{\Gamma(D \to \pi\pi)}{\Gamma_{tot}}.
\end{equation}
\begin{table}[h]
\caption{{\small The branching ratios of $D \to \pi\pi$ obtained in the naive
factorization approach and compared with the experimental
results.}}
\begin{center}
\begin{tabular}{|c|c|c|c|} \hline
Decay mode & Br (Theory) & Br (Experiment)\\
\hline
$D^+ \to \pi^+\pi^0$ &$3.1 \times 10^{-3}$ & $(2.5\pm 0.7)\times 10^{-3}$ \\
\hline
$D^0 \to \pi^+\pi^-$ &$2.48 \times 10^{-3}$ &$(1.52 \pm 0.09) \times
10^{-3}$\\
\hline
$D^0 \to \pi^0\pi^0$ &$1.0 \times 10^{-7}$  & $(8.4 \pm 2.2) \times
10^{-4}$\\
\hline
\end{tabular}
\end{center}
\end{table}
A comparison of the branching ratios of the naive factorization result with
the experimental data is presented in Table 1, where the second column
gives the pure factorization result. One can notice
that the results are not in agreement with
the experimental data, especially the second and third decay modes.

\section{The one particle exchange method for FSI}

As we have seen above, the experimental results for the branching ratios of
$D^0 \to \pi^+\pi^-$ and $D^0\to \pi^0\pi^0$ are in disagreement with
the calculation from the naive factorization approach. The reason is that
the physical picture of naive factorization is too simple,
nonperturbative strong interaction is restricted in single
hadrons, or between the initial and final hadrons which share the same
spectator quark. If the mass of the initial particle is large, such as
the case of $B$ meson decay, the effect of nonperturbative strong
interaction between the final hadrons most probably is small because
the momentum transfer is large. However, in the case of $D$ meson,
its mass is not so large. The energy scale of $D$ decays is not
very high. Nonperturbative effect may give large contribution.
Because there exist many resonances near the mass scale of $D$
meson, it is possible that nonperturbative interaction is propagated
by these resonance states, such as, $K^*(892)$, $K^*(1430)$,
$f_0(1710)$, etc.

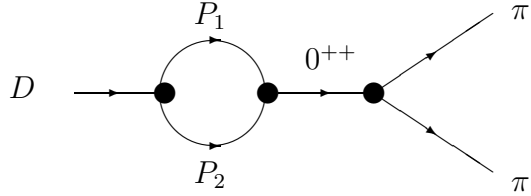
\begin{figure}[h]
\begin{center}
\begin{picture}(250,90)
      \put(35,50){\line(1,0){30}}
      \put(69,50){\circle*{8}}
      \put(87,50){\circle{80}}
      \put(108,50){\circle*{8}}
      \put(112,50){\line(1,0){38}}
      \put(148,50){\circle*{8}}
      \put(148,50){\line(3,-2){45}}
      \put(148,50){\line(3,2){45}}

   \put(50,50){\vector(3,0){2}}
   \put(88,70){\vector(3,0){2}}
   \put(88,30){\vector(3,0){2}}
   \put(130,50){\vector(3,0){2}}
   \put(170,64.5){\vector(3,2){2}}
   \put(170,35.5){\vector(3,-2){2}}

      \put(10,48){$D$}
      \put(80,16){$P_2$}
      \put(80,76){$P_1$}
      \put(122,59){$0^{++}$}
      \put(200,78){$\pi$}
      \put(200,13){$\pi$}
   \end{picture}
 \end{center}
 \caption{{\small s-channel contributions to final-state interaction in $D\to
     \pi \pi $ due to one particle  exchange.}}
 \label{schan}
\end{figure}

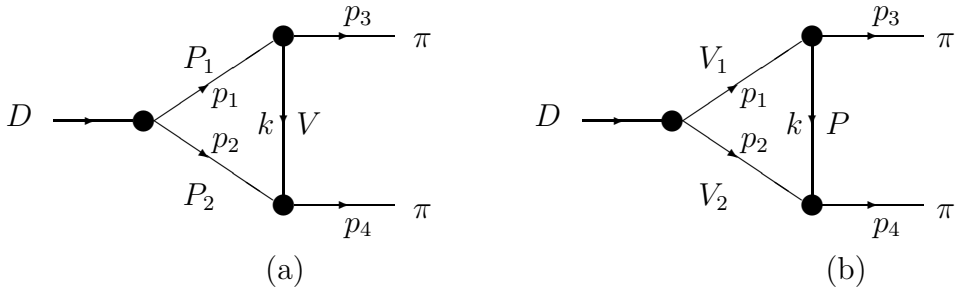
\begin{figure}[h]
\begin{center}
\begin{picture}(430,160)
\put(28,65){\line(1,0){30}}
\put(62,65){\circle*{8}}
\put(66,65){\line(3,2) {45}}
\put(66,65){\line(3,-2){45}}
\put(115,33){\circle*{8}}
\put(115,97){\circle*{8}}
\put(115,37){\line(0,1){56}}
\put(119,33){\line(1,0){38}}
\put(119,97){\line(1,0){38}}

\put(42,65){\vector(3,0){2}}
\put(85,52){\vector(3,-2){2}}
\put(85,77.5){\vector(3,2){2}}
\put(115,65){\vector(0,-2){2}}
\put(138,97){\vector(3,0){2}}
\put(138,33){\vector(3,0){2}}

\put(10,63){$D$}
\put(77,33){$P_2$}
\put(77,85){$P_1$}
\put(120,60){$V$}
\put(88,72){$p_1$}
\put(88,55){$p_2$}
\put(105,60){$k$}
\put(164,93){$\pi$}
\put(164,28){$\pi$}
\put(138,103){$p_3$}
\put(138,23){$p_4$}
\put(108,5){(a)}

\put(228,65){\line(1,0){30}}
\put(262,65){\circle*{8}}
\put(266,65){\line(3,2) {45}}
\put(266,65){\line(3,-2){45}}
\put(315,33){\circle*{8}}
\put(315,97){\circle*{8}}
\put(315,37){\line(0,1){56}}
\put(319,33){\line(1,0){38}}
\put(319,97){\line(1,0){38}}

 \put(242,65){\vector(3,0){2}}
 \put(285,52){\vector(3,-2){2}}
 \put(285,77.5){\vector(3,2){2}}
 \put(315,65){\vector(0,-2){2}}
 \put(338,97){\vector(3,0){2}}
 \put(338,33){\vector(3,0){2}}

  \put(210,63){$D$}
  \put(272,33){$V_2$}
  \put(272,85){$V_1$}
  \put(320,60){$P$}
  \put(288,72){$p_1$}
  \put(288,55){$p_2$}
  \put(305,60){$k$}
  \put(362,93){$\pi$}
  \put(362,28){$\pi$}
  \put(338,103){$p_3$}
  \put(338,23){$p_4$}
  \put(320,5){(b)}
\end{picture}
\end{center}
\caption{{\small t-channel contributions to final-state interaction in $D\to
 \pi \pi $ due to one particle  exchange.
 (a) Exchange a single vector meson, (b) Exchange a single
pseudoscalar meson}}
 \label{tchan}
\end{figure}

The diagrams of these nonperturbative rescattering effects can be
depicted in Figs.\ref{schan} and \ref{tchan}.
The first part $D \to P_1P_2$ or $D \to V_1V_2$ represents the direct
decay where the decay amplitudes can be obtained by using naive
factorization method. The second part represents rescattering process
where the effective hadronic couplings are needed in numerical
calculation, which can be extracted from experimental data on the
relevant resonance decays.

Fig.\ref{schan} is the $s$-channel contribution to the final
state interaction. Here $P_1$ and $P_2$ are the intermediate pseudoscalar
mesons. The resonance state has the quantum number $J^{PC}=0^{++}$
derived from the final state particles $\pi\pi$. From
Particle Data Group~\cite{PDG},
one can only choose $f_0(1710)$ as the resonance state
which evaluates the $s$-channel contribution.

Fig.\ref{tchan} shows the $t$-channel contribution to the final
state interaction. $P_1$, $P_2$ and $V_1$, $V_2$ are the intermediate
states. They rescatter into the final state $\pi\pi$ by exchanging
one resonance state $V$ or $P$. In this paper the intermediate states
are treated to be on their mass shell, because their off-shell contribution
can be attributed to the quark level. We assume the on-shell contribution
dominates in the final state interaction. The exchanged resonances are
treated as a virtual particle. Their propagators are taken as
Breit-Wigner form
\begin{equation}
\frac{i}{k^2-m^2+im\Gamma_{tot}}, \nonumber
\end{equation}
where $\Gamma_{tot}$ is the total decay width of the exchanged resonance.

To the lowest order, the effective couplings of $f_0$ to $PP$ and $VV$
can be taken as the form
\begin{eqnarray}
L_I & =& g_{fPP} \phi^+\phi f,\\
L_I & =& g_{fVV} A_\mu A^\mu f,
\end{eqnarray}
where $\phi$ is the pseudoscalar field, $A_\mu$ the vector field. Then
the decay amplitudes of $f_0\to PP$ and $VV$ are
\begin{eqnarray}
T_{fPP} & =& g_{fPP},\label{t1}\\
T_{fVV} & =& g_{fVV} \epsilon_\mu\epsilon^{\mu}.
\end{eqnarray}
The coupling constants $g_{fPP}$ and $g_{fVV}$ can
be extracted from the measured branching fractions of $f_0\to PP$
and $VV$ decays, respectively \cite{PDG}.  Because $f_0\to VV$ decays
have not been detected yet, we
assume their couplings are small. We do not consider the
intermediate vector meson contributions in $s$-channel in this paper.

For the $t$-channel contribution, the concerned effective vertex
is $VPP$, which can be related to the $V$ decay amplitude. Explicitly
the amplitude of $V\to PP$ can be written as
\begin{equation}
T_{VPP}=g_{VPP}\;\epsilon\cdot (p_1-p_2),
\label{t2}
\end{equation}
where $p_1$ and $p_2$ are the four-momentum of the two pseudoscalars,
respectively. To extract $g_{fPP}$ and $g_{VPP}$ from experiment,
one should square eqs.(\ref{t1}) and (\ref{t2}) to get the decay widths
\begin{eqnarray}
\Gamma(f\to PP)&=& \frac{1}{8\pi}\mid g_{fPP}\mid^2
           \frac{\mid\vec{p}\mid}{m_f^2},\nonumber\\
\Gamma(V\to PP)&=& \frac{1}{3}\frac{1}{8\pi}\mid g_{VPP}\mid^2
         \left[m_V^2-2m_1^2-2m_2^2+\frac{(m_1^2-m_2^2)^2}{m_V^2}\right]
          \frac{\mid\vec{p}\mid}{m_V^2},
\label{couple}
\end{eqnarray}

\begin{figure}
\begin{center}
\begin{picture}(150,60)

\put(28,35){\line(1,0){38}}
\put(65,35){\circle*{8}}
\put(66,35){\line(3,2) {45}}
\put(66,35){\line(3,-2){45}}

\put(42,35){\vector(3,0){2}}
\put(85,22){\vector(3,-2){2}}
\put(85,47.5){\vector(3,2){2}}

\put(10,33){$K^*$}
\put(116,0){$\pi$}
\put(116,62){$K$}

\end{picture}
\end{center}
\caption{{\small The effective coupling vertex on the hadronic
level}}
\label{vertex1}
\end{figure}
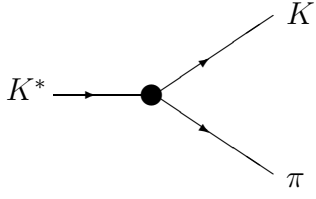
\begin{figure}
\begin{center}
\begin{picture}(150,100)

\put(36,65){\line(3,0){30}}
\put(36,40){\line(3,0){30}}
\put(66,65){\line(3,2) {45}}
\put(66,40){\line(3,-2){45}}
\put(73,52){\line(3,2) {45}}
\put(73,52){\line(3,-2){45}}

\GlueArc(63,67)(10,195,375){2}{8}
\GlueArc(63,38)(10,-20,170){2}{8}
\Gluon(66,65)(73,52){2}{3}
\Gluon(66,40)(73,52){2}{3}
\GlueArc(73,60)(15,160,300){2}{15}

\put(42,65){\vector(3,0){2}}
\put(42,40){\vector(-3,0){2}}

\put(100,88){\vector(3,2){2}}
\put(105,73.5){\vector(-3,-2){2}}

\put(105,30.5){\vector(3,-2){2}}
\put(99,18){\vector(-3,2){2}}

\put(-8,50){$K_0^*$}
\put(23,63){$d$}
\put(23,38){$\bar s$}
\put(100,4){$\bar s$}
\put(100,96){$d$}
\put(110,34){$u$}
\put(110,66){$\bar u$}

\put(125,6){$K^0$}
\put(128,91){$\pi$}

\end{picture}
\end{center}
\caption{{\small The effective coupling vertex on the quark
level}}
\label{vertex2}
\end{figure}
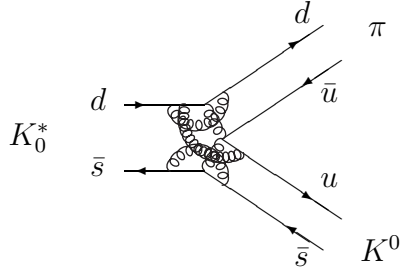

\noindent where $m_1$ and $m_2$ are the masses of the two final particles $PP$,
respectively, and $\mid\vec{p}\mid$ is the momentum of one of the final
particle $P$ in the rest frame of $V$ or $f$. From the above equations,
one can see that only the magnitudes of the effective couplings
$\mid g_{fPP}\mid $ and $\mid g_{VPP}\mid $ can be extracted from
experiment. If there is any phase factor for the effective coupling,
it would be dropped. Actually it is quite possible that there are
imaginary phases for the effective couplings. As an example, let us
see the effective coupling of $g_{K^*K\pi}$ shown in Fig.\ref{vertex1},
which relevant to the process $K^* \to K\pi$. On the quark level,
the effective vertex can be depicted as Fig.\ref{vertex2}, which should
be controlled by nonperturbative QCD. From this figure one can see that
it is resonable that a strong phase could appear in the effective
coupling, which is contributed by strong interaction.
Therefore we can introduce a strong phase for each
hadronic effective
coupling. In the succeeding part of this paper, the symbol $g$
will only be used to represent the magnitude of the relevant
effective coupling. The total one should be $g e^{i \theta}$, where
$\theta$ is the strong phase coming from Fig.\ref{vertex2}. For example,
the effective couplings will be written in the form of
$g_{fPP}e^{i\theta_{fPP}}$ and $g_{VPP}e^{i\theta_{VPP}}$.

The decay amplitude of the $s$-channel final state interaction
can be calculated from Fig.\ref{schan}
\begin{equation}
  A^{FSI}_s=\frac12 \int \frac{d^3 \vec p_1}{(2\pi)^3
2E_1} \int \frac{d^3 \vec p_2}{(2\pi)^3 2E_2} (2\pi)^4
\delta^4(p_D-p_1-p_2) \; A(D \to
P_1P_2)\frac{i\;g_1\;g_2\;e^{i(\theta_1+\theta_2)}}{k^2-m^2+im\Gamma_{tot}},
\end{equation}
where $p_1$ and $p_2$ represent the four-momenta of the pseudoscalar $P_1$
and $P_2$, the amplitude $A(D \to P_1P_2)$ is the direct decay
amplitude.  The effective coupling constants $g_1$ and $g_2$
should be $g_{fPP}$ or $g_{VPP}$ which can be obtained by
comparing eq.(\ref{couple}) with experimental data.
By performing integrals, we can obtain
\begin{equation}
  A^{FSI}_s=\frac{1}{8\pi m_D} |\vec p_1| A(D \to P_1P_2)
  \frac{i\;g_1\;g_2\;e^{i(\theta_1+\theta_2)}}{k^2-m^2+im\Gamma_{tot}}.
\end{equation}

The $t$-channel contribution via exchanging a vector
meson (Fig.\ref{tchan}(a) ) is
\begin{eqnarray}
  A^{FSI}_{t,V}&=&\frac12 \int \frac{d^3 \vec p_1}{(2\pi)^3 2E_1}
          \int\frac{d^3 \vec p_2}{(2\pi)^3 2E_2} (2\pi)^4
 \delta^4(p_D-p_1-p_2) A(D \to P_1P_2)\nonumber\\
  & &\hspace*{1cm}\times g_1 \; \epsilon_\lambda \cdot (p_1+p_3)
  \frac{i\;e^{i(\theta_1+\theta_2)}}{k^2-m^2+im\Gamma_{tot}} \;
 F(k^2)^2 \;g_2 \; \epsilon_\lambda\cdot(p_2+p_4),
\label{fsisk}
\end{eqnarray}
where
$F(k^2)=(\Lambda^2-m^2)/(\Lambda^2-k^2)$
is the
form factor which is introduced to compensate the off-shell effect of the
exchanged particle at the vertices \cite{Fk}. In the numerical calculation we take
$\Lambda =0.5 GeV$. We choose the lightest
resonance state as the exchanged particle that gives rise to the largest
contribution to the decay amplitude.

We can furthermore obtain
\begin{equation}
  A^{FSI}_{t,V}=\int^1_{-1}  \frac{d(\cos\theta)}{16\pi m_D} |\vec p_1|
   A(D \to P_1P_2) \;g_1 \frac{i\;e^{i(\theta_1+\theta_2)}}
  {k^2-m^2+im\Gamma_{tot}} \;F(k^2)^2\;g_2\;H ,
\label{fsitv}
\end{equation}
where
\begin{eqnarray}
  H&=&-\left[m_D^2 -\frac12(m_1^2+m_2^2+m_3^2+m_4^2)
     +(|\vec p_1||\vec p_4|+ |\vec p_2||\vec p_3|)\cos\theta
     +E_1E_4+E_2E_3 \right]\nonumber\\
   & & -\frac{1}{m_V^2}(m_1^2-m_3^2)(m_2^2-m_4^2).
\end{eqnarray}
The $t$-channel contribution by exchanging a pseudoscalar meson
(Fig.\ref{tchan}(b) )is
\begin{eqnarray}
  A^{FSI}_{t,P}&=&\frac12 \int \frac{d^3 \vec p_1}{(2\pi)^3 2E_1}
          \int\frac{d^3 \vec p_2}{(2\pi)^3 2E_2} (2\pi)^4
 \delta^4(p_D-p_1-p_2) \sum_{\lambda_1,\lambda_2}A(D \to V_1V_2)\nonumber\\
  & &\hspace*{1cm} \times g_1 \; \epsilon_{\lambda_1} \cdot (p_3-k)
  \frac{i\;e^{i(\theta_1+\theta_2)}}{k^2-m^2+im\Gamma_{tot}} \; F(k^2)^2
    \;g_2 \; \epsilon_{\lambda_2}\cdot(k+p_4),
\end{eqnarray}
and we obtain
\begin{eqnarray}
 A^{FSI}_{t,P}=\int^1_{-1} \frac{d(\cos\theta)}{16\pi m_D} |\vec p_1| \;
    \frac{i\;e^{i(\theta_1+\theta_2)}}{k^2-m^2+im\Gamma_{tot}}
      X \; g_1 \;g_2 \; F(k^2)^2\;(-H_1+H_2),
\label{fsitp}
\end{eqnarray}
where
\begin{eqnarray}
  H_1&=&4im_{V_1}f_{V_1}(m_D+m_2)
A_1\left[\frac12(m_D^2-m_3^2-m_4^2)\right.\nonumber\\
    &&\hspace*{4cm} -\frac{1}{m_1^2} ( E_1E_3-|\vec p_1||\vec
p_3|\cos\theta)
          ( E_1E_4+|\vec p_1||\vec p_4|\cos\theta)\nonumber\\
  &&\hspace*{4cm} -\frac{1}{m_2^2}( E_2E_4-|\vec p_2||\vec
      p_4|\cos\theta) ( E_2E_3+|\vec p_2||\vec p_3|\cos\theta)\nonumber\\
& &\left.+\frac{1}{2m_1^2m_2^2}( m_D^2-m_1^2-m_2^2 )
 ( E_1E_3-|\vec p_1||\vec p_3|\cos\theta )(E_2E_4-|\vec p_2||\vec
p_4|\cos\theta) \right],
\end{eqnarray}
\begin{eqnarray}
H_2&=&\frac{8im_{V_1}f_{V_1}}{(m_D+m_2)}A_2
     \left[ E_2E_3+|\vec p_2||\vec p_3|\cos\theta -
  \frac{1}{2m_1^2}(m_D^2-m_1^2-m_2^2)(E_1E_3-|\vec p_1||\vec
p_3|\cos\theta)\right]\nonumber\\
& &\hspace*{1cm}\left[ E_1E_4+|\vec p_1||\vec p_4|\cos\theta -
  \frac{1}{2m_2^2}(m_D^2-m_1^2-m_2^2)( E_2E_4-|\vec p_2||\vec
p_4|\cos\theta ) \right],
\end{eqnarray}
and $X$ represents the relevant direct decay amplitude of $D$ decaying
to the intermediate vector pair $V_1$ and $V_2$ divided by
$\langle V_1 |(V-A)_\mu |0\rangle \langle V_2|(V-A)^\mu|D\rangle$,
$$
X\equiv \frac{A(D\to V_1 V_2)}
  {\langle V_1 |(V-A)_\mu |0\rangle \langle
  V_2|(V-A)^\mu|D\rangle}.
$$

\section{Numerical calculation and  discussions}

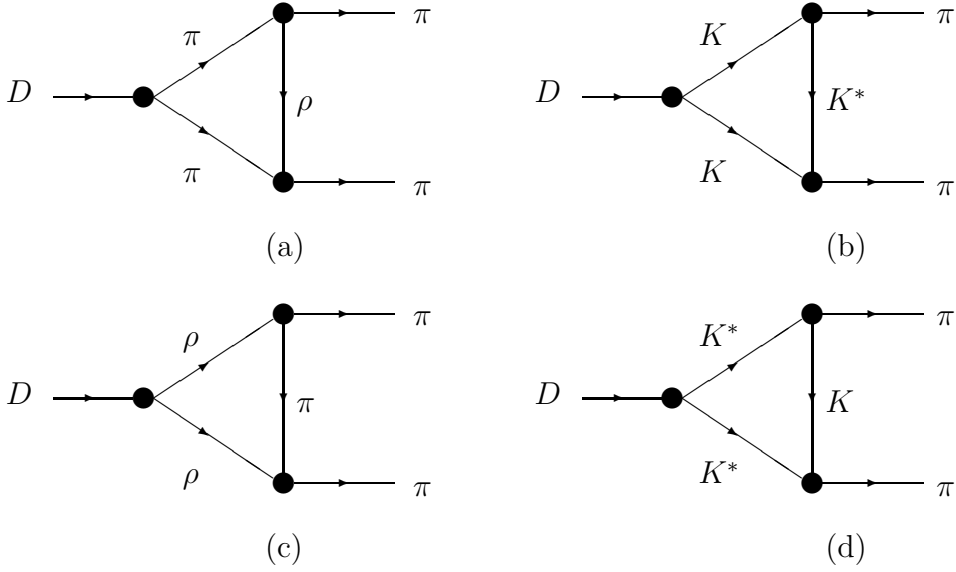
\begin{figure}
\begin{center}
\begin{picture}(430,100)
\put(28,65){\line(1,0){30}}
\put(62,65){\circle*{8}}
\put(66,65){\line(3,2) {45}}
\put(66,65){\line(3,-2){45}}
\put(115,33){\circle*{8}}
\put(115,97){\circle*{8}}
\put(115,37){\line(0,1){56}}
\put(119,33){\line(1,0){38}}
\put(119,97){\line(1,0){38}}

\put(42,65){\vector(3,0){2}}
\put(85,52){\vector(3,-2){2}}
\put(85,77.5){\vector(3,2){2}}
\put(115,65){\vector(0,-2){2}}
\put(138,97){\vector(3,0){2}}
\put(138,33){\vector(3,0){2}}

\put(10,63){$D$}
\put(77,33){$\pi$}
\put(77,85){$\pi$}
\put(120,60){$\rho$}
\put(164,93){$\pi$}
\put(164,28){$\pi$}
\put(108,5){(a)}

\put(228,65){\line(1,0){30}}
\put(262,65){\circle*{8}}
\put(266,65){\line(3,2) {45}}
\put(266,65){\line(3,-2){45}}
\put(315,33){\circle*{8}}
\put(315,97){\circle*{8}}
\put(315,37){\line(0,1){56}}
\put(319,33){\line(1,0){38}}
\put(319,97){\line(1,0){38}}

 \put(242,65){\vector(3,0){2}}
 \put(285,52){\vector(3,-2){2}}
 \put(285,77.5){\vector(3,2){2}}
 \put(315,65){\vector(0,-2){2}}
 \put(338,97){\vector(3,0){2}}
 \put(338,33){\vector(3,0){2}}

  \put(210,63){$D$}
  \put(272,33){$K$}
  \put(272,85){$K$}
  \put(320,60){$K^*$}
  \put(362,93){$\pi$}
  \put(362,28){$\pi$}
  \put(320,5){(b)}
\end{picture}
\end{center}

\begin{center}
\begin{picture}(430,100)
\put(28,65){\line(1,0){30}}
\put(62,65){\circle*{8}}
\put(66,65){\line(3,2) {45}}
\put(66,65){\line(3,-2){45}}
\put(115,33){\circle*{8}}
\put(115,97){\circle*{8}}
\put(115,37){\line(0,1){56}}
\put(119,33){\line(1,0){38}}
\put(119,97){\line(1,0){38}}

\put(42,65){\vector(3,0){2}}
\put(85,52){\vector(3,-2){2}}
\put(85,77.5){\vector(3,2){2}}
\put(115,65){\vector(0,-2){2}}
\put(138,97){\vector(3,0){2}}
\put(138,33){\vector(3,0){2}}

\put(10,63){$D$}
\put(77,33){$\rho$}
\put(77,85){$\rho$}
\put(120,60){$\pi$}
\put(164,93){$\pi$}
\put(164,28){$\pi$}
\put(108,5){(c)}

\put(228,65){\line(1,0){30}}
\put(262,65){\circle*{8}}
\put(266,65){\line(3,2) {45}}
\put(266,65){\line(3,-2){45}}
\put(315,33){\circle*{8}}
\put(315,97){\circle*{8}}
\put(315,37){\line(0,1){56}}
\put(319,33){\line(1,0){38}}
\put(319,97){\line(1,0){38}}

 \put(242,65){\vector(3,0){2}}
 \put(285,52){\vector(3,-2){2}}
 \put(285,77.5){\vector(3,2){2}}
 \put(315,65){\vector(0,-2){2}}
 \put(338,97){\vector(3,0){2}}
 \put(338,33){\vector(3,0){2}}

  \put(210,63){$D$}
  \put(272,33){$K^*$}
  \put(272,85){$K^*$}
  \put(320,60){$K$}
  \put(362,93){$\pi$}
  \put(362,28){$\pi$}
  \put(320,5){(d)}
\end{picture}
\end{center}

\caption{{\small Intermediate states in rescattering process
for $D\to\pi\pi$ decays}} \label{scatter}
\end{figure}

In general, every decay channel should be analysed to see whether it can
rescatter into $\pi\pi$ state, and how large the contribution is
if it can. Here for $D\to\pi\pi$ decays, the rescattering processes
$D\to\pi\pi\to\pi\pi$, $D\to KK\to\pi\pi$, and $D\to\rho\rho\to\pi\pi$,
$D\to K^*K^*\to\pi\pi$ give the largest contributions, because these
intermediate states have the largest couplings with the final state
pion and the exchanged meson state shown in Fig.\ref{scatter}.
The contribution of each diagram in Fig.\ref{scatter} should not only
be calculated via eqs.(\ref{fsitv}) and (\ref{fsitp}), there
is but also an isospin factor for each diagram which should be multiplied to
the calculation of eqs.(\ref{fsitv}) or (\ref{fsitp}). The isospin factor
should be considered in such a way that, at first, consider all the possible
isospin structure for each diagram in Fig.\ref{scatter} and draw all
the possible sub-diagrams on the quark level. One diagram in Fig.\ref{scatter}
may amount to several diagrams on quark level. Second, write down the isospin
factor for each sub-diagram. For example, the $u\bar{u}$ component
in one final meson $\pi^0$ contributes an isospin factor $\frac{1}{\sqrt{2}}$,
and the $d\bar{d}$ component contributes $-\frac{1}{\sqrt{2}}$. For the
intermediate state $\pi^0$, the factor $\frac{1}{\sqrt{2}}$ and
$-\frac{1}{\sqrt{2}}$ should be dropped, otherwise, the isospin relation
between these three channels $D^+\to \pi^+\pi^0$,
$D^0\to \pi^+\pi^-$ and $D^0\to \pi^0\pi^0$ would be violated. Third, sum
the contributions of all the possible diagrams on the quark level together
to get the isospin factor for each diagram on the hadronic level.

We study these three channels of $D\to \pi\pi$ decays: $D^+\to \pi^+\pi^0$,
$D^0\to \pi^+\pi^-$ and $D^0\to \pi^0\pi^0$. In $D^+\to \pi^+\pi^0$, the
isospin factors for Fig.\ref{scatter}(b) and (d) are zero since the contributions
of the sub-diagrams on the quark level
cancel each other. The rescattering contribution in $D^+\to \pi^+\pi^0$ only
depends on $g_{\rho\pi\pi}e^{i\theta\rho\pi\pi}$. In $D^0\to \pi^0\pi^0$ decay, however,
the FSI contribution only depends on $g_{K^*K\pi}e^{i\theta_{K^*K\pi}}$, because the isospin
factors of Fig.\ref{scatter}(a) and (c) are zero for the sub-diagrams of them
cancel each other. The FSI contribution in $D^0\to \pi^+\pi^-$ decay depends
on both $g_{\rho\pi\pi}e^{i\theta\rho\pi\pi}$ and $g_{K^*K\pi}e^{i\theta_{K^*K\pi}}$. The
numerical results of the branching ratios of these three $D\to \pi\pi$
decays including both the direct decay and the rescattering effects are:
\begin{eqnarray}
Br(D^+\to \pi^+\pi^0)&=&1.68\times 10^{10}\mid 4.29\times 10^{-7}i
  -(6.84\times 10^{-7}+8.69\times 10^{-8}i)e^{i2\theta_{\rho\pi\pi}}\mid ^2,
   \nonumber\\
Br(D^0\to \pi^+\pi^-)&=&6.61\times 10^{9}\mid -6.13\times 10^{-7}i
  -(5.41\times 10^{-7}+2.43\times 10^{-8}i)e^{i2\theta_{K^*K\pi}}+
   \nonumber\\
  &&~~~~~~~~~(9.67\times 10^{-7}+1.23\times 10^{-7}i)
e^{i2\theta_{\rho\pi\pi}}\mid ^2, \label{bnr}
   \\
Br(D^0\to \pi^0\pi^0)&=&6.61\times 10^{9}\mid -3.89\times 10^{-9}i
  -(3.82\times 10^{-7}+1.72\times 10^{-8}i)e^{i2\theta_{K^*K\pi}}\mid ^2.
   \nonumber
\end{eqnarray}

The above equations satisfy the isospin relation
\begin{equation}
\frac{1}{\sqrt 2}A(D^0 \to \pi^+\pi^-) - A(D^0 \to \pi^0\pi^0) = -A(D^+ \to
\pi^+\pi^0).\nonumber
\end{equation}
In order to get~(\ref{bnr}), we have used
 eq.(\ref{couple}) and the center
value of the measured decay width of $\rho\to\pi\pi$
and $K^*\to K\pi$ \cite{PDG} to
obtain $g_{\rho\pi\pi}=6.0$, $g_{K^*K\pi}=4.6$.
While using the measured value
of $f_0(1710)\to KK$ and $f_0(1710)\to \pi\pi$ decays \cite{PDG},
one can get $g_{fKK}=1.6$ and $g_{f\pi\pi}=0.29$.
Comparing the value of $g_{fPP}$ and $g_{VPP}$, we
can see that the amplitude of s-channel contribution to
FSI is at least 40 times ($1.6\times 0.29/4.6^2$)
 smaller than t-channel
contribution. Therefore we can drop the s-channel
contribution in our numerical analysis. The other
input parameters used in the numerical calculation are:
1) the form factors, $F^{D\pi}(0)=0.692$, $F^{DK}(0)=0.762$,
$A_1^{D\rho}(0)=0.775$, $A_2^{D\rho}(0)=0.923$, $A_1^{DK}(0)=0.880$,
$A_2^{DK}(0)=1.147$ \cite{BSW}; 2) the decay constants,
$f_{\pi}=0.133GeV$, $f_K=0.158GeV$, $f_D=0.205GeV$,
$f_{\rho}=0.2GeV$, and $f_{K^*}=0.2GeV$; 3)
$\Lambda=0.5 GeV$.

\begin{figure}[htbp]
\begin{center}
\scalebox{0.9}{
  \epsfig{file=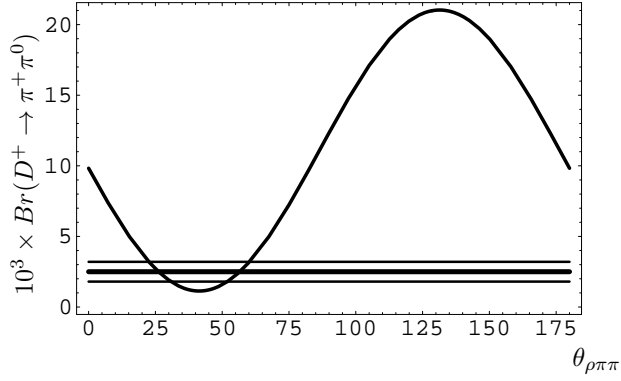}
    \begin{picture}(0,0)(60,60)
      \put(-190,80){ \small{\rotatebox{90}{$10^3\times Br(D^+\to\pi^+\pi^0)$}} }
      \put(50,50){\small{$\theta_{\rho\pi\pi}$}}
    \end{picture} }
\end{center}
\caption{{\small The branching ratio of $D^+\to\pi^+\pi^0$ vs.
     $\theta_{\rho\pi\pi}$. The horizontal lines are the
     centeral value of experimental data and error bars.}}
\label{figp0}
\end{figure}

\begin{figure}[htbp]
\begin{center}
\scalebox{0.9}{
  \epsfig{file=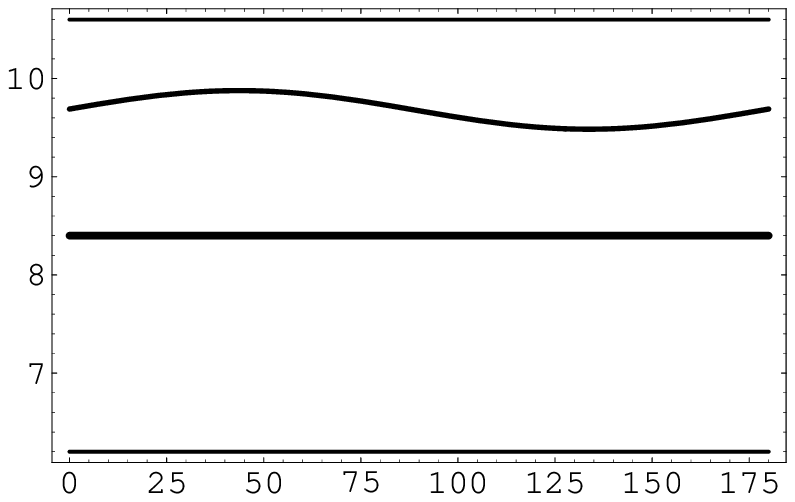}
    \begin{picture}(0,0)(60,60)
      \put(-190,80){ \small{\rotatebox{90}{$10^4\times Br(D^0\to\pi^0\pi^0)$}} }
      \put(50,50){\small{$\theta_{K^*K\pi}$}}
    \end{picture} }
\end{center}
\caption{{\small The branching ratio of $D^0\to\pi^0\pi^0$ vs.
     $\theta_{K^*K\pi}$. The horizontal lines are the
     centeral value of experimental data and error bars.}}
\label{fig00}
\end{figure}

\begin{figure}[htbp]
\begin{center}
\scalebox{0.9}{
  \epsfig{file=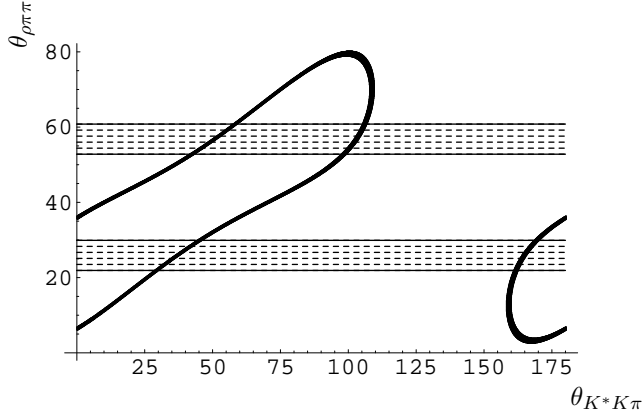}
    \begin{picture}(0,0)(60,60)
      \put(-190,200){ \small{\rotatebox{90}{$\theta_{\rho\pi\pi}$}} }
      \put(50,50){\small{$\theta_{K^*K\pi}$}}
    \end{picture} }
\end{center}
\caption{{\small The allowed region of $\theta_{\rho\pi\pi}$
and $\theta_{K^*K\pi}$
by experimental data. The solid belts are from
$D^0\to \pi^+\pi^-$, and the dashed belts
allowed by $D^+\to \pi^+\pi^0$. The overlap is
allowed by both data.}}
\label{figpn}
\end{figure}

The strong phase $\theta_{\rho\pi\pi}$ and $\theta_{K^*K\pi}$ can
not be known from any other existing data at present, they
are treated as free parameters. Fig.\ref{figp0} and \ref{fig00}
show the branching ratios of $D^+\to\pi^+\pi^0$ and
$D^0\to\pi^0\pi^0$ changing with the strong phase
$\theta_{\rho\pi\pi}$ and $\theta_{K^*K\pi}$, respectively. The
ranges $(22.6^\circ, 30.8^\circ)$ and $(51.9^\circ, 60.2^\circ)$
for $\theta_{\rho\pi\pi}$ are allowed by the measured branching
ratio of $D^+\to\pi^+\pi^0$, while the whole range of
$\theta_{K^*K\pi}$ is allowed by $D^0\to\pi^0\pi^0$.
Fig.\ref{figpn} shows the allowed region of
$\theta_{\rho\pi\pi}$ and $\theta_{K^*K\pi}$ by the three decay
modes $D^+\to\pi^+\pi^0$, $D^0\to\pi^+\pi^-$ and
$D^0\to\pi^0\pi^0$. In the overlap regions in Fig.\ref{figpn} all
the three decay modes are in agreement with the experimental data.
However, if $SU(3)$ symmetry is kept with small violation, the
relation $\theta_{\rho\pi\pi}\simeq \theta_{K^*K\pi}$ should be
satisfied. Considering this relation, only one of the four overlap
regions in Fig.\ref{figpn} shall be allowed,  where
$51.9^\circ < \theta_{\rho\pi\pi} < 60.2^\circ$ and
$40.9^\circ < \theta_{K^*K\pi} < 60.7^\circ$. As an example, Table
\ref{tabbr2} shows the branching ratios of $D^+\to\pi^+\pi^0$,
$D^0\to\pi^+\pi^-$ and $D^0\to\pi^0\pi^0$ by taking one sample
point in the overlap region of Fig.\ref{figpn}
$(\theta_{\rho\pi\pi}, \theta_{K^*K\pi})=(57.3^\circ, 51.0^\circ
)$, where the $SU(3)$ symmetry violation is in the order of a few
degrees. Column `Factorization' is for the branching ratio
predicted in naive factorization approach. Column `Factorization +
FSI' is for the branching ratio of naive factorization including
the final state interaction. The contributions of final state
rescattering effects are large, which can improve the predictions
of naive factorization to be consistent with the experimental
data. The strong phases introduced for the effective hadronic
couplings $g_{\rho\pi\pi}$ and $g_{K^*K\pi}$ are important for
explaining the experimental data, otherwise, it is quite difficult
to get the correct results for the three decay modes at the same
time by varying other input parameters.

\begin{table}
\caption{{\small The branching ratios of $D \to \pi\pi$.}}
\begin{center}
\begin{tabular}{|c|c|c|c|} \hline
Decay mode & Factorization & Factorization + FSI & Experiment\\
\hline
$D^+ \to \pi^+\pi^0$ &$3.1 \times 10^{-3}$&
$2.63\times 10^{-3}$&$(2.5\pm 0.7)\times 10^{-3}$ \\
\hline
$D^0 \to \pi^+\pi^-$ &$2.48 \times 10^{-3}$&
$1.57\times 10^{-3}$ &$(1.52 \pm 0.09) \times 10^{-3}$\\
\hline
$D^0 \to \pi^0\pi^0$ &$1.0 \times 10^{-7}$ &
$9.9\times 10^{-4}$& $(8.4 \pm 2.2) \times 10^{-4}$\\
\hline
\end{tabular}
\end{center}
\label{tabbr2}
\end{table}

The parameter $\Lambda$ in the off-shellness compensating function
$F(k^2)$ introduced in eq.(\ref{fsisk}) takes the value $0.5GeV$
in this calculation, while in Ref.\cite{ope1,Fk} the value takes
$\Lambda=1.2\sim 2.0 GeV$, that is because it is quite possible that
$\Lambda$ is not an universal parameter. We assume that $\Lambda$
should be near the masses of the mesons involved in the effective
coupling. The reaction studied in Ref.\cite{ope1,Fk}
is $\bar{P}P\to \phi\pi$, where
the parameter $\Lambda$ should be near the mass of
$\phi$ meson $m_{\phi}=1.02 GeV$. Therefore its value can be
taken to be $\Lambda=1.2\sim 2.0 GeV$. While in the decay process
studied in this paper, $K$, $\pi$, $K^*$ and $\rho$ are involved.
Therefore the parameter $\Lambda$ can take about $0.5GeV$, which is located in
the range of the masses of these mesons. However, it is still
necessary to study the off-shell properties of the hadronic effective
couplings in a direct nonperturbative way to check the shape of
$F(k^2)$ used in eq.(\ref{fsisk}), because the numerical results
of FSI rescattering effects calculated in this model
are sensitively dependent on the off-shell properties of the
hadronic couplings, or specifically to say, the shape of
the off-shell compensating function
$F(k^2)$. To show this dependence, we give the results of the
three decay branching ratios in Table \ref{tabbr3} by varying the
value of the parameter $\Lambda$. It shows that the branching
ratios are very sensitive to the variation of $\Lambda$. Certainly
there are also many other free parameters, such as the form
factors, some meson decay constants which have not yet been well
determined in experiment. So the allowed value of the strong
phases $\theta_{\rho\pi\pi}$ and $\theta_{K^*K\pi}$ may heavily
depends on these parameters. Therefore Fig. \ref{figpn} shall not
be viewed as a stringent constraint on the strong phases. It only shows
the possibility to accommodate the three $D\to \pi\pi$ decay modes consistently
in this model. Certainly to completely understand final state interaction,
more experimental data and more detailed theoretical works are needed.

\begin{table}
\caption{{\small The dependence of the branching ratios of
    $D \to \pi\pi$ on different values of $\Lambda$, where
    $(\theta_{\rho\pi\pi}, \theta_{K^*K\pi})=(57.3^\circ, 51.0^\circ)$.}}
\begin{center}
\begin{tabular}{|c|c|c|c|} \hline
$\Lambda({\mathrm{GeV}})$ & $D^+ \to \pi^+\pi^0$ &$D^0 \to \pi^+\pi^-$ & $D^0 \to \pi^0\pi^0$\\
\hline
0.45 & $9.23 \times 10^{-3}$& $2.57\times 10^{-3}$& $2.2\times 10^{-3}$ \\
\hline
0.5 & $2.63 \times 10^{-3}$& $1.57\times 10^{-3}$ & $9.9 \times 10^{-4}$\\
\hline
0.55 &$9.48 \times 10^{-4}$ & $1.39\times 10^{-3}$& $4.24 \times 10^{-4}$\\
\hline
\end{tabular}
\end{center}
\label{tabbr3}
\end{table}

{\bf Summary} We have studied three $D\to \pi\pi$ decay modes. The total
decay amplitude includes direct weak decays and final state rescattering
effects. The direct weak decays are calculated in factorization
approach, and the final state interaction effects are studied in
one-particle-exchange method. The prediction of naive factorization
is far from the experimental data. After including the contribution
of final state interaction, the theoretical prediction can accommodate
the experimental data.

\vspace{5mm}

\noindent {\large{\bf Acknowledgement}}\vspace{0.3cm}

\noindent This work is supported in part by National Natural Science
Foundation of China. M. Yang thanks the partial support of the Research Fund
for Returned Overseas Chinese Scholars. M. Ablikim is grateful to
Scientific Research Foundation for  Returned Scholars of State Education
Ministry of China.

\end{document}